\documentclass[10pt, preprint]{aastex}

\usepackage{siunitx}
\usepackage[square,sort,comma,numbers]{natbib}
\usepackage{subcaption} 
\usepackage{listings}
\usepackage{xcolor}
\usepackage{textgreek}
\definecolor{codegreen}{rgb}{0,0.6,0}
\definecolor{codegray}{rgb}{0.5,0.5,0.5}
\definecolor{codepurple}{rgb}{0.58,0,0.82}
\definecolor{backcolour}{rgb}{0.95,0.95,0.92}
\usepackage{tikz}
\usetikzlibrary{shapes.geometric, arrows, positioning}
\usepackage{footmisc}
\tikzstyle{block} = [rectangle, draw, text centered, minimum height=3em, minimum width=3cm]
\tikzstyle{line} = [draw, -latex']
\lstdefinestyle{mystyle}{
    backgroundcolor=\color{backcolour},   
    commentstyle=\color{codegreen},
    keywordstyle=\color{magenta},
    numberstyle=\tiny\color{codegray},
    stringstyle=\color{codepurple},
    basicstyle=\ttfamily\footnotesize,
    breakatwhitespace=false,         
    breaklines=true,                 
    captionpos=b,                    
    keepspaces=true,                 
    numbers=left,                    
    numbersep=5pt,                  
    showspaces=false,                
    showstringspaces=false,
    showtabs=false,                  
    tabsize=2
}
\usepackage{listings}
\usepackage{float}
\usepackage{graphicx}
\usepackage{amsmath}
\usepackage[toc,page]{appendix}
\usepackage[utf8]{inputenc}
\usepackage{hyperref}
\hypersetup{
    colorlinks=true,
    linkcolor=blue,
    filecolor=magenta,      
    urlcolor=blue,
    citecolor=blue,
}
\usepackage{booktabs}
\renewcommand{\arraystretch}{1}

\title{Development and Characterization of a Novel BaTiO$_3$-Based Material for Medium Temperature Applications}

\author{
Weitian Chen\altaffilmark{1}, 
Songyang Bai\altaffilmark{1}, 
Zihan Gao\altaffilmark{2},
Kaiheng Ding\altaffilmark{3}\\
}
\altaffiltext{1}{Department of Physics, University of Toronto, weitian.chen@mail.utoronto.ca(W.C.), songyang.bai@mail.utoronto.ca(S.B.)}
\altaffiltext{2}{Department of Physics, University of Alberta, zgao13@ualberta.ca}
\altaffiltext{3}{Department of Engineering, Virginia Polytechnic Institute and State University}
\begin{document}
\begin{abstract}

Positive temperature coefficient (PTC) materials are extensively utilized in self-regulating temperature applications. Nonetheless, their applicability is typically constrained to low-temperature ranges, rendering them ineffective in medium temperature environments. This study presents a methodology for the fabrication of an innovative PTC material operational at approximately 353 ℃, with a thorough investigation of its Curie temperature and resistivity properties. The material formulation incorporates 4 wt\% carbon black (CB), 0.5 wt\% NBT, and 5 wt\% DOP into a BaTiO$_3$-based matrix. The empirical findings reveal that this material exhibits a notably high PTC strength of 5.8 and a comparatively low resistivity of 590 $\Omega \cdot$cm at room temperature. Furthermore, the material demonstrated excellent repeatability in PTC strength after thirty cycles of heating and cooling near the Curie temperature. Consequently, this PTC material is deemed highly effective for applications in cold environments, notably for the preheating and initiation of aircraft engines and auxiliary power units (APUs).

\end{abstract}

\maketitle

\section{Introduction}

The relentless global pursuit of advanced materials capable of withstanding extreme environments has catalyzed the development of substances exhibiting a positive temperature coefficient (PTC) effect. These materials are distinguished by their cost-effectiveness, lightweight nature, pronounced flexibility, and inherent self-regulation, thereby presenting considerable potential for applications in thermal management systems. A quintessential feature of PTC materials is their pronounced temperature sensitivity, wherein resistivity experiences a significant upsurge as the temperature escalates, a phenomenon occurring at what is referred to as the Curie temperature \cite{reference3}. Owing to these distinctive properties, PTC materials find extensive utility within the electrical industry, particularly in the development of self-limiting fuses \cite{reference1} and sensors \cite{reference2,reference4}, in addition to their application in adaptive thermal control systems, including self-regulating heating cables \cite{reference4,reference5,reference6,reference7}. The research endeavors of Cheng, Xu, and their colleagues have elucidated the application of PTC composites for temperature-adaptive control, a development that obviates the need for conventional temperature control electronics, thereby enabling a reduction in the mass or volume of thermal management systems. Moreover, PTC composites can be seamlessly integrated with proportional-integral-derivative (PID) components to enhance the precision and accuracy of thermal regulation systems \cite{reference8,reference9}. The Curie temperature of the majority of extant PTC materials typically spans a range of 40 to 500°C \cite{reference10,reference11,reference12}. Nonetheless, PTC composites with elevated Curie temperatures frequently exhibit diminished hardness, thus rendering them unsuitable for applications necessitating repeated mechanical endurance \cite{reference13}. In a seminal study, Cheng et al. developed a composite material characterized by a resistivity of 103 \textOmega$\cdot$cm at a Curie temperature of 340°C, with an associated PTC strength of approximately 3 \cite{reference14}. Subsequent investigations have revealed that by lowering the Curie temperature to 18°C and inducing a first-order phase transition, the PTC strength can be modulated and reduced to 2. \cite{reference15,reference16}, the meticulous development of PTC materials that balance appropriate resistivity with low mechanical strength has been a focal point of research. Cheng et al. have identified an optimal PTC material, which exhibits a strength of 5.0 and a resistivity of approximately 13,600 \textOmega$\cdot$cm \cite{reference17}. Investigations into amorphous PS/SPE-MWCNT composites have revealed a significantly higher resistivity of 105 \textOmega$\cdot$cm \cite{reference18}. It is noteworthy that increased resistivity can induce higher excitation voltages, thereby potentially escalating the risk of electric shock. Within the scope of this study, a PTC material with a Curie temperature of 353°C was successfully synthesized. This material is distinguished by its exceptional strength and low resistivity at ambient temperature, while maintaining superior resistivity and mechanical robustness at elevated temperatures. The study also explored the impact of variations in raw material dosages on resistivity at the Curie temperature and PTC strength. Furthermore, the repeatability of the optimized formulation was rigorously evaluated. The interplay between the Curie temperature and the melting point of PTC composites was scrutinized using differential scanning calorimetry (DSC). Additionally, the distribution of conductive particles and their consequent influence on the PTC composites’ properties were meticulously analyzed via scanning electron microscopy (SEM).

\section{Materials and Methodology}
\label{sec:material}
\subsection{Materials for the compostie}
Acetylene carbon black (CB) is provided by Graphene Market, and TiBaO$_3$ is made in this research using BaCO$_3$, Nb$_2$O$_5$, Sb$_2$O$_3$ and TiO$_2$ provided by MSE supplies LLC and Santech LTD.. NBT, also known as Na$_{0.5}$Bi$_{0.5}$TiO$_3$ are composed in this research. The material used are powders of Na$_2$CO$_3$, Bi$_2$O$_3$, and TiO$_2$
. Bi$_2$O$_3$ is provided by Santech LTD., TiO$_2$ is provided by Ingridient Depot LLC., tacetone is provided by Lab Alley LTD. and the xylene and DOP (Dioctyl phthalate) were prepared by Sigma-Aldrich.

\subsection{Composite process of PTC material}
\begin{figure}[H]
    \centering
    \includegraphics[width=0.8\linewidth]{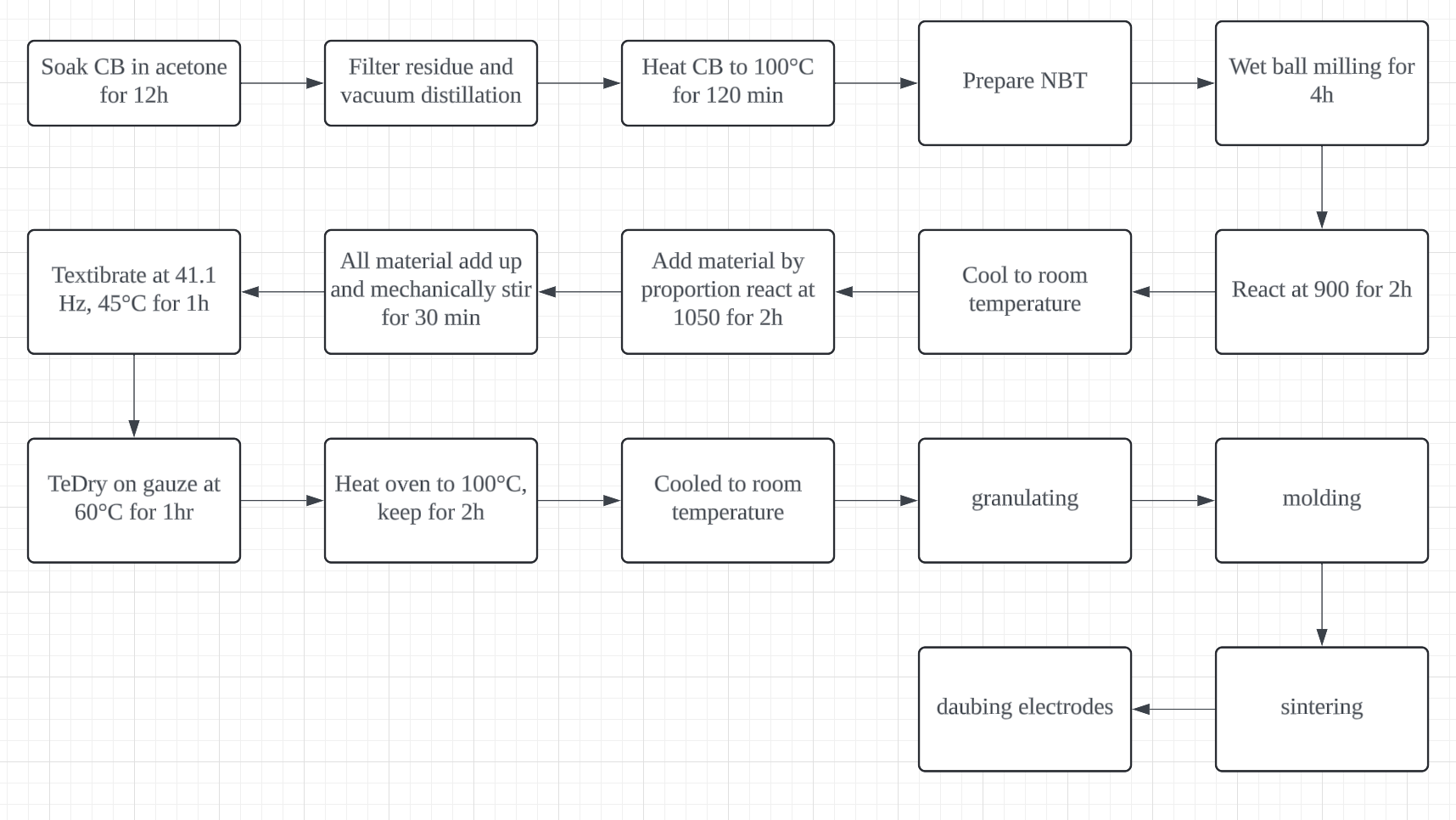}
    \caption{Flow Chart for this experiment}
    \label{Fig1}
\end{figure}

Acetylene carbon black (CB) was immersed in \( \text{C}_3\text{H}_6\text{O} \) liquid for 12 hours to remove any other matters adsorbed on its surface. After filtering out the residue, \( \text{C}_3\text{H}_6\text{O} \) liquid was eliminated by vacuum distillation. Subsequently, the carbon black(CB) was heated to 100°C for 120 minutes to remove substances from its surface. The next step involved preparing NBT, where high-purity Na$_2$CO$_3$, Bi$_2$O$_3$, and TiO$_2$ powders were mixed using wet ball milling for 4 hours. This mixture was then reacted at 900°C for 2 hours and allowed to cool to room temperature. BaTiO$_3$ was synthesized by adding appropriate amounts of BaCO$_3$, Nb$_2$O$_5$, Sb$_2$O$_3$, and TiO$_2$, followed by another wet ball milling for 4 hours, drying at 120°C in an oven for 3 hours, and reacting at 1050°C for 2 hours. Subsequently, CB powder, liquid NBT, and DOP were incorporated into the BaTiO$_3$ based on the ratio outlined in Table 1. This mixture was mechanically stirred for 30 minutes, then subjected to ultrasonic vibration at 41.1 Hz and 45°C for 1 hour. The mixture was then placed on a 50 mm thick gauze and dried in an oven at 60°C for 1 hour. The oven temperature was set to 100°C, maintaining it for 2 hours. The final product was cooled to room temperature in a fan drying box. Afterward, the compound was granulated, molded, and sintered, preparing it for electrode application. The process is illustrated in the flow chart shown in Figure \ref{Fig1}.
\begin{table}[!ht]
\centering
\begin{tabular}{>{\centering\arraybackslash}p{1cm} >{\centering\arraybackslash}p{2cm} >{\centering\arraybackslash}p{2cm} >{\centering\arraybackslash}p{2cm} >{\centering\arraybackslash}p{1cm} >{\centering\arraybackslash}p{2cm} >{\centering\arraybackslash}p{2cm} >{\centering\arraybackslash}p{2cm}}
\toprule
\textbf{Number} & \textbf{CB} & \textbf{NBT/ BaTiO$_3$} & \textbf{DOP} & \textbf{Number} & \textbf{CB} & \textbf{NBT/BaTiO$_3$} & \textbf{ DOP} \\
\midrule
1 & 1\% & 1:3 & -- & 4A & 4\% & 1:4 & -- \\
2 & 2\% & 1:3 & -- & 4B & 4\% & 1:3.5 & -- \\
3 & 3\% & 1:3 & -- & 4C & 4\% & 1:3 & -- \\
4 & 4\% & 1:3 & -- & 4D & 4\% & 1:2.5 & -- \\
5 & 5\% & 1:3 & -- & 4E & 4\% & 1:2 & -- \\
6 & 6\% & 1:3 & -- & A4 & 4\% & 1:3 & -- \\
7 & 7\% & 1:3 & -- & B4 & 4\% & 1:3 & 5\% \\
8 & 8\% & 1:3 & -- & C4 & 4\% & 1:3 & 10\% \\
9 & 9\% & 1:3 & -- & D4 & 4\% & 1:3 & 15\% \\
10 & 10\% & 1:3 & -- & E4 & 4\% & 1:3 & 20\% \\
\bottomrule
\end{tabular}
\caption{Ratios of CB, NBT/BaTiO$_3$, and xylene in the experiment}
\end{table}
\subsection{Method for measurement}
Following the test, the resistivity (\( \rho \)) of the materials was determined using the formula \( \rho = \frac{dV}{dI} \times \frac{A}{L} \), where \( \frac{dV}{dI} \) represents the differential resistance, \( A \) is the cross-sectional area, and \( L \) is the length of the sample. This approach allows for a more accurate representation of the material's resistive properties, especially under varying electrical loads. The melting behavior of the PTC composites was meticulously analyzed using differential scanning calorimetry (DSC) in a nitrogen atmosphere, applying a controlled heating rate of 5℃/min. This technique enables a detailed study of thermal transitions, such as melting points and crystallization behaviors, which are essential for understanding the thermal stability and processing characteristics of the materials. To gain insights into the microstructural characteristics of the samples, a scanning electron microscope (SEM) was employed. This high-resolution imaging technique provides detailed images of the sample surface, allowing for the analysis of the morphology, size, and distribution of grains, as well as the distribution and morphology of conductive particles within the composites. Such analysis is crucial for correlating the microstructure with the electrical and thermal properties of the PTC materials, thereby guiding the optimization of material formulations and processing conditions.

\section{Result and Discussion}\label{sec:result}
\subsection{Positive temperature coefficient Effect at Different content at room temperature}
The investigation commenced with an analysis of how varying carbon black (CB) content affects the material properties. As demonstrated in Figures 1 and 2, the resistivity of composite samples, measured at 25\(^\circ\)C, changes in response to different CB mass fractions. The results indicate a percolation threshold at around 4\%. For CB contents less than 2\%, the resistivity at room temperature (25\(^\circ\)C) exhibits a gradual decline, eventually stabilizing near 110 $\Omega\cdot$cm. This stability suggests that the material acts as an insulator under these conditions. When the CB content ranges between 2\% and 4\%, there is a sharp decrease in room temperature resistivity, dropping from 10$^8$ $\Omega\cdot$cm to 10$^2$ $\Omega\cdot$cm. This significant reduction indicates a shift in the material’s properties from insulating to semiconducting. Interestingly, when the CB content exceeds 4\%, the resistivity at room temperature shows little to no further change, indicating a saturation point in the conductive network within the composite.
\begin{figure}[h!]
    \centering
    \begin{subfigure}[b]{0.49\textwidth}
        \centering
        \includegraphics[width=\textwidth]{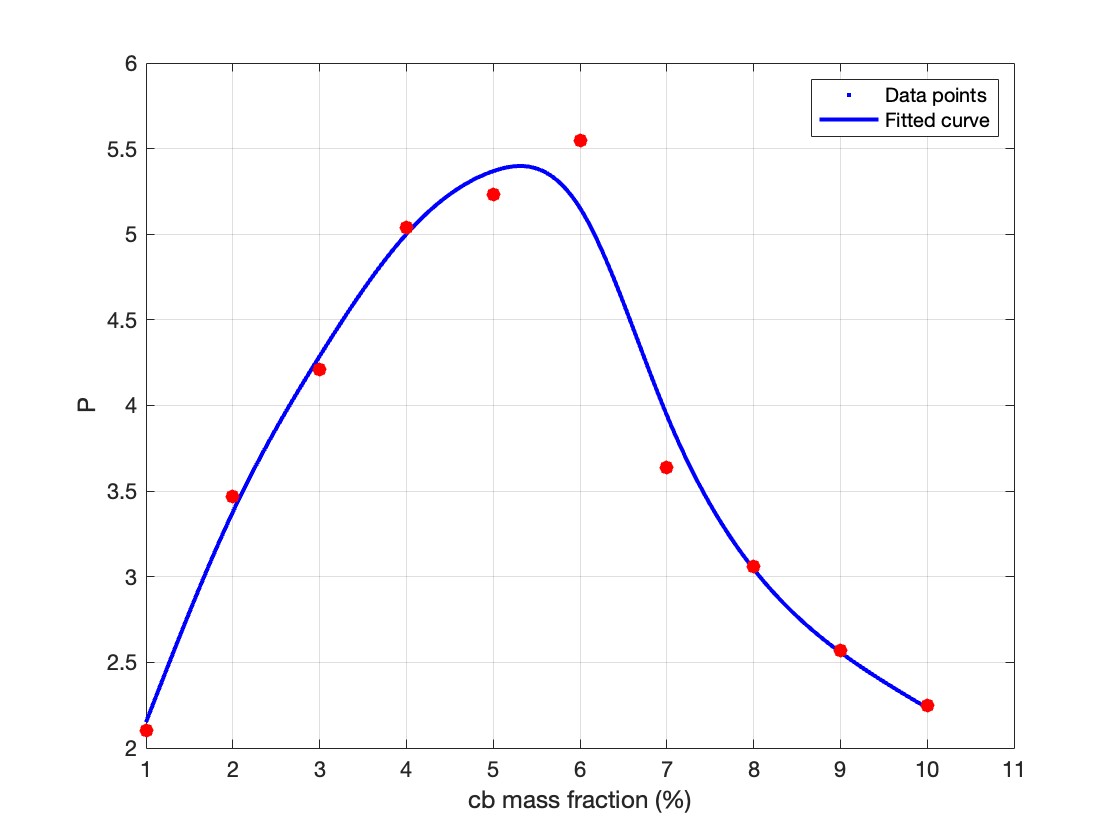}
        \caption{Intensity - CB mass fraction curve for 10 different samples}
        \label{fig:figure2a}
    \end{subfigure}
    \hfill
    \begin{subfigure}[b]{0.49\textwidth}
        \centering
        \includegraphics[width=\textwidth]{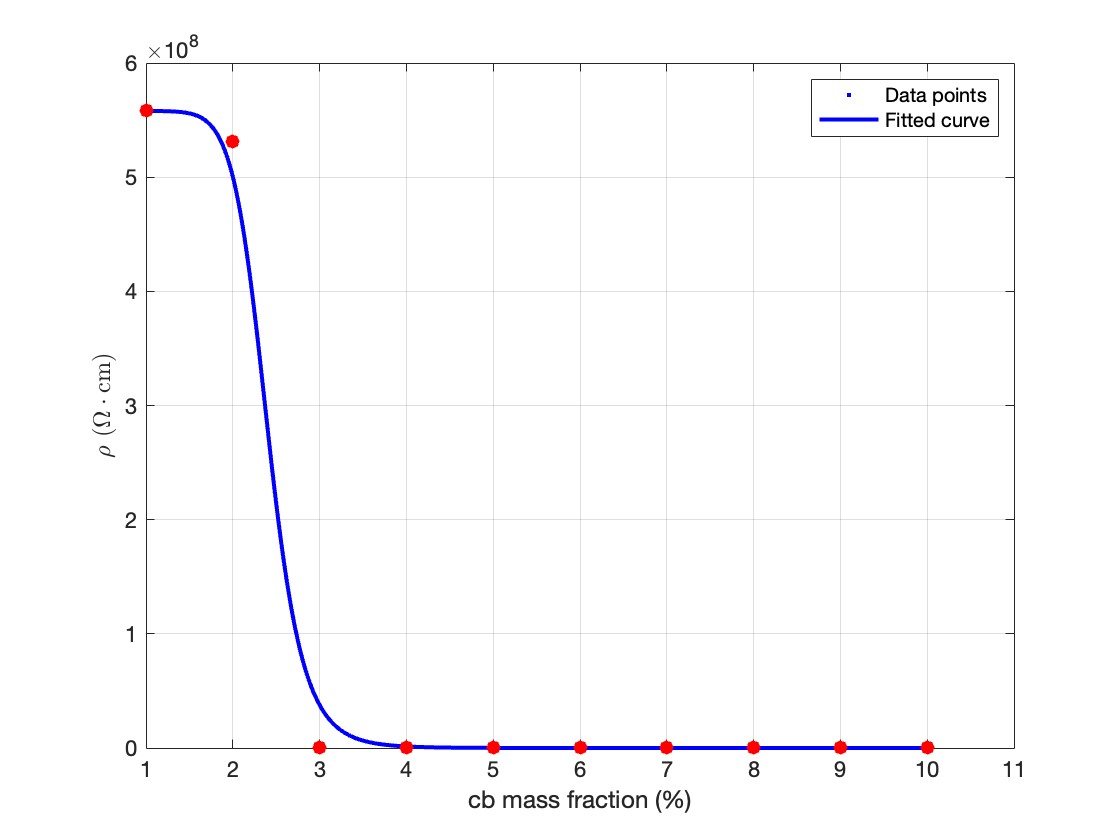}
        \caption{Resistivity - CB mass fraction curve for 10 different samples}
        \label{fig:figure2b}
    \end{subfigure}
    \hfill\\
    \begin{subfigure}[b]{0.49\textwidth}
        \centering
        \includegraphics[width=\textwidth]{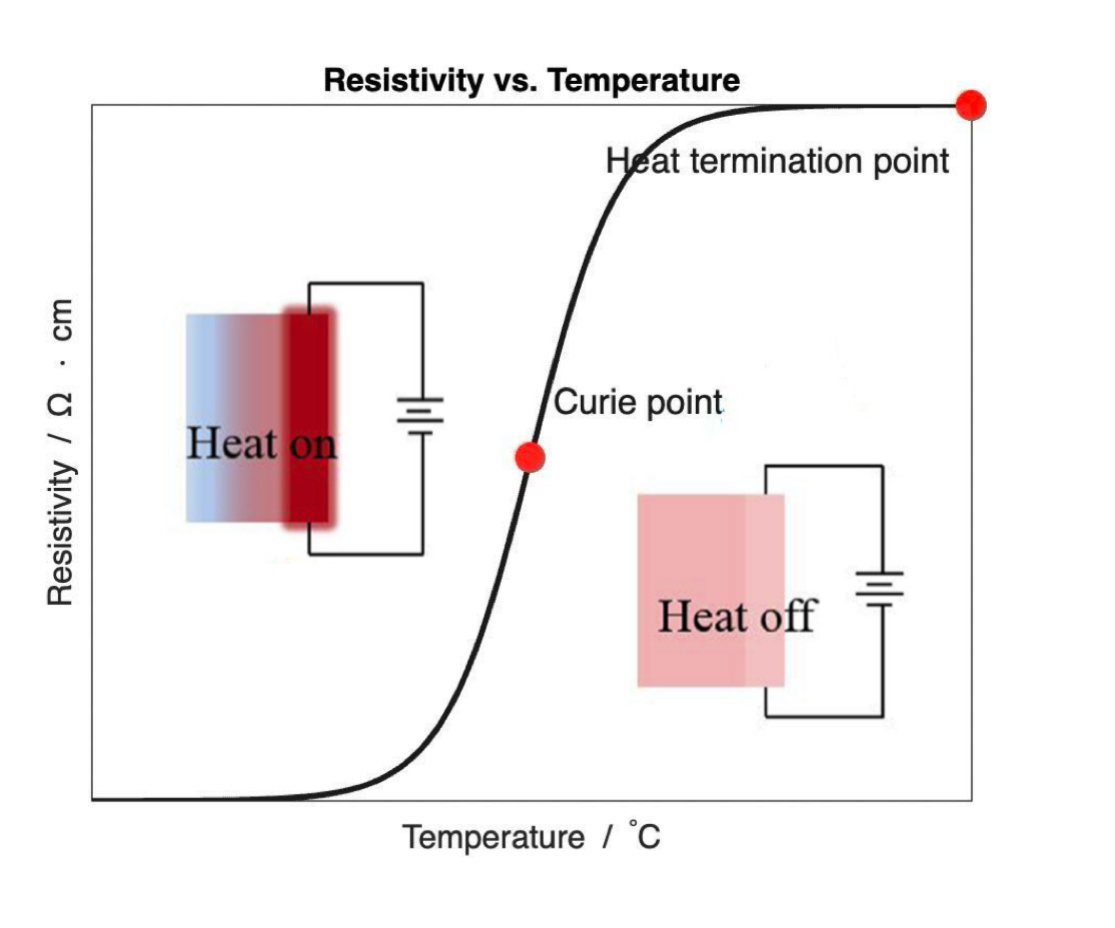}
        \caption{Example curie point}
        \label{fig:figure2b}
    \end{subfigure}
    \caption{a): The Intensity(P) vs. CB mass fraction curve for 10 different samples, b): The Resistivity(ρ) vs. CB mass fraction curve for 10 different samples, c) example curie point in the curve}
    \label{fig: figure 2}
\end{figure}

\begin{figure}[H]
    \centering
    \includegraphics[width=0.6\linewidth]{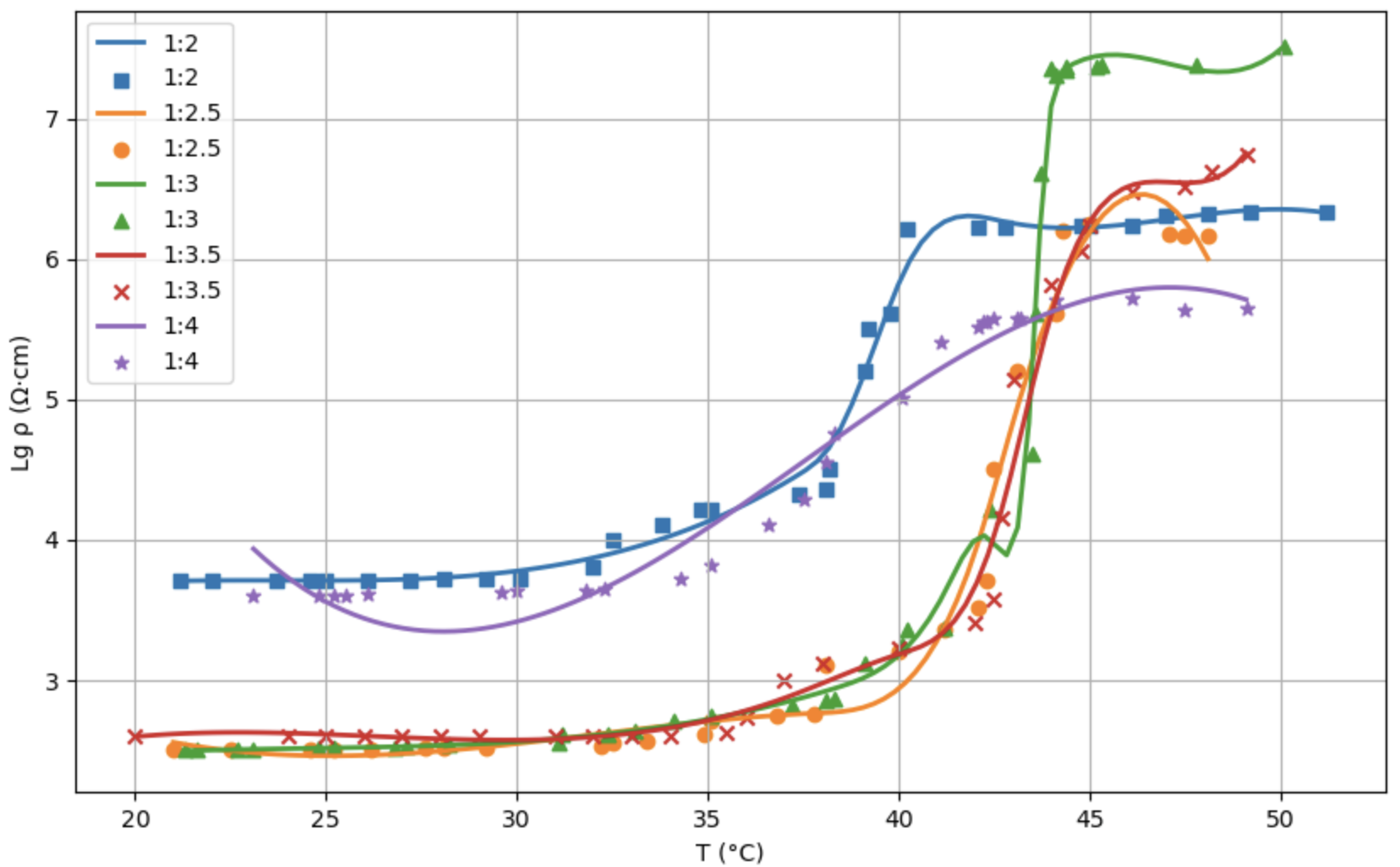}
    \caption{Logrithemtic resistivity($\rho$)-temperature(°C) curves at varient content of NBT/TiBaO\(_3\) for composite 4A-4E}
    \label{fig:figure 3}
\end{figure}

The research first focused on the impact of carbon black (CB) mass fraction on the PTC intensity (\(P\)), which is defined as the logarithmic ratio of maximum resistivity (\(\rho_{\text{max}}\)) to minimum resistivity (\(\rho_{\text{min}}\)) across a given temperature range, as shown in Figure 2c. Through careful analysis, it was observed that the CB mass fraction near the percolation threshold, specifically at 4\%, provided an optimal balance, where the resistivity at room temperature (\(\rho\)) was approximately 590 \(\Omega\cdot\)cm and the PTC intensity (\(P\)) reached 5.7, indicating the most effective configuration for thermal control applications (see Figure \ref{fig: figure 2}a). Building on this finding, the CB content was kept constant at the optimal level of 4\%, while the ratio of NBT (Sodium Bismuth Titanate) to TiBaO\(_3\) was varied. Importantly, the combined content of NBT and TiBaO\(_3\) was maintained at 92\% to ensure the integrity of the composite material's properties. The temperature-resistivity behavior of samples 4A to 4E, under these conditions, is illustrated in Figure \ref{fig:figure 3}. To further clarify the findings, the resistivity at room temperature and the corresponding PTC strength derived from these samples were compiled and are presented in Table \ref{tab:table2}. Additionally, fitting equations, which are essential for understanding the relationship between the material composition and its thermal behavior, are provided in Section \ref{sec: Appendix}. It is critical to note that any deviation from the optimized CB content or the carefully controlled NBT to TiBaO\(_3\) ratio results in either insufficient or excessive resistivity, which in turn diminishes the PTC intensity. This highlights the delicate balance required in the material composition to achieve the desired thermal control characteristics.

\begin{table}[H]
    \centering
    \caption{Resistivity and PTC intensity of model 4A--4E.}
    \label{tab:table2}
    \scalebox{1.2}{
    \begin{tabular}{cccccc}
        \toprule
        \textbf{Model number} & \textbf{4A} & \textbf{4B} & \textbf{4C} & \textbf{4D} & \textbf{4E} \\
        \midrule
        NBT/TiBaO\(_3\) ratio & 1:4 & 1:3.5 & 1:3 & 1:2.5 & 1:2 \\
        25\(^\circ\)C resistivity (Ω$\cdot$cm) & 4750 & 340 & 590 & 430 & 7090 \\
        PTC intensity P & 2.6 & 3.72 & 5.8 & 4.23 & 2.64 \\
        \bottomrule
    \end{tabular}}
\end{table}

\begin{figure}[h!]
    \centering
    \begin{subfigure}[b]{0.5\textwidth}
        \centering
        \includegraphics[width=\textwidth]{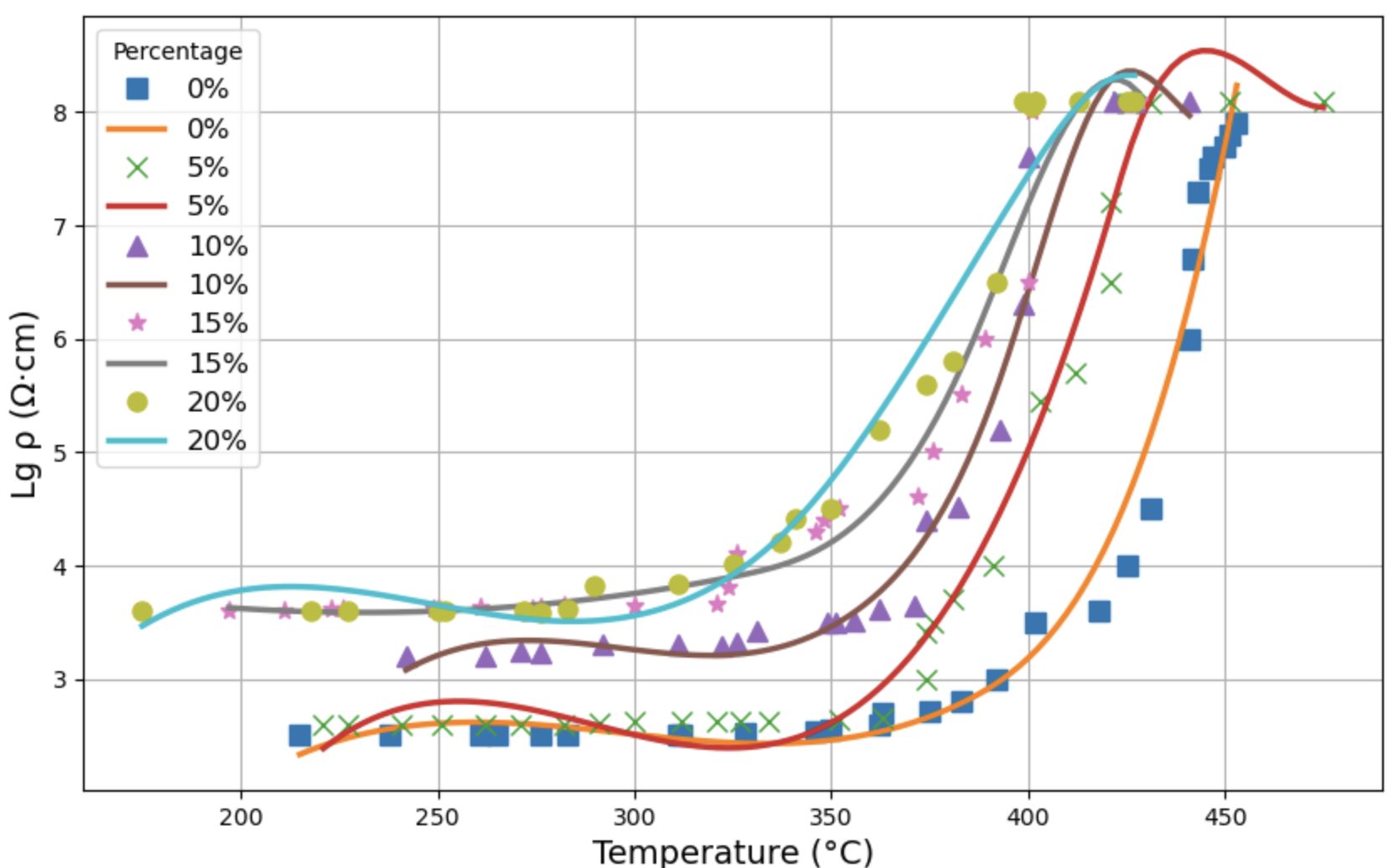}
        \label{fig:figure4a}
    \end{subfigure}
    \hfill
    \begin{subfigure}[b]{0.49\textwidth}
        \centering
        \includegraphics[width=\textwidth]{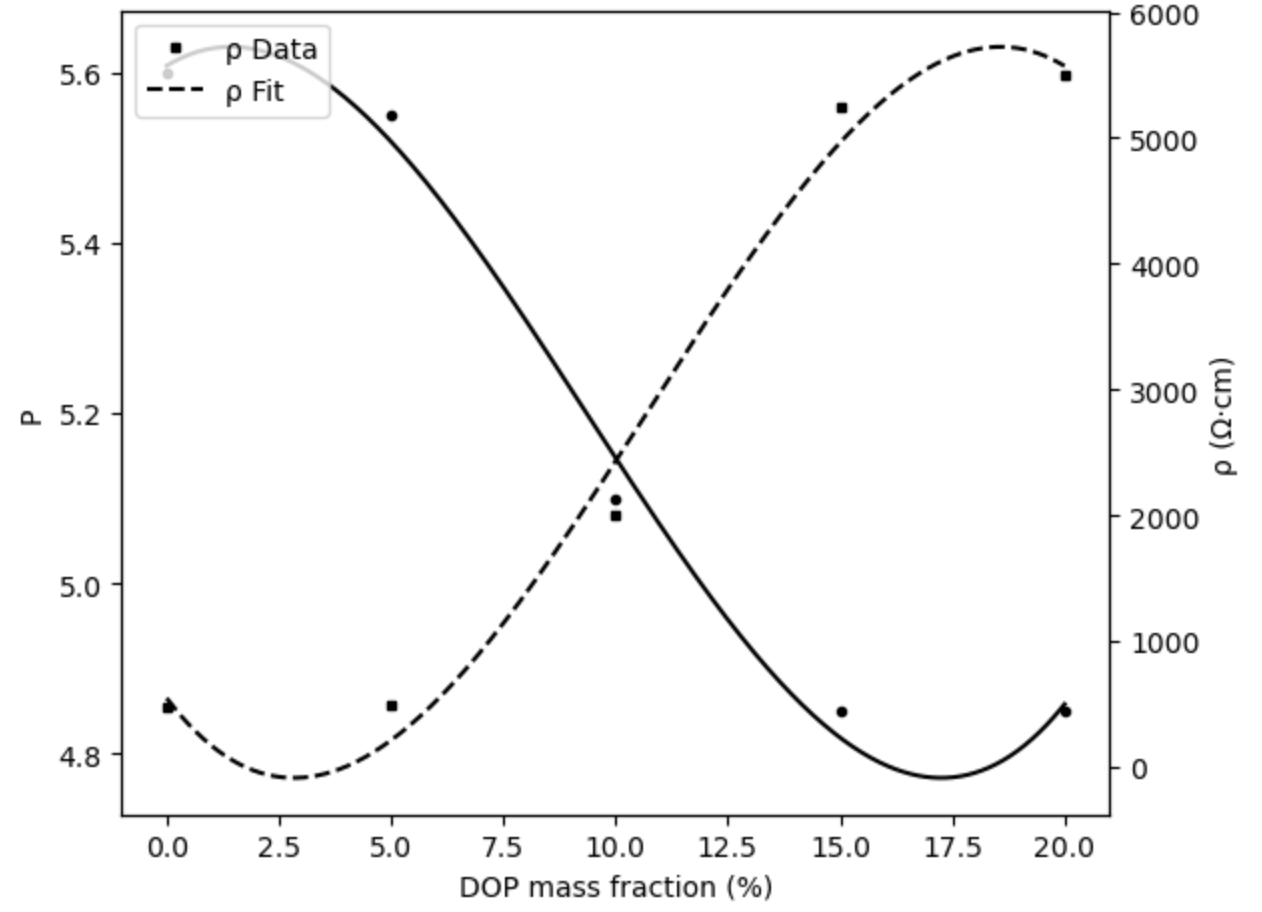}
        \label{fig:figure4b}
    \end{subfigure}
    \caption{Impact of Dioctyl Phthalate (DOP) Concentration on Material Characteristics. (a) Logarithmic analysis of material property variations across different DOP levels. (b) Correlation between PTC intensity and resistivity at varying DOP Concentrations.}
    \label{fig: figure 4}
\end{figure}

In analyzing the temperature-dependent resistivity behaviors of samples 4A through 4E, a distinct trend emerges. The logarithmic resistivity temperature curves for 4A and 4E display a more confined vertical range, indicating that their resistivity changes less drastically with temperature increases compared to other samples. This contrasts with samples 4B and 4D, where the resistivity increases are somewhat more pronounced, though still less significant than that observed in sample 4C. As shown in Table \ref{tab:table2}, sample 4C stands out with the highest PTC intensity (\(P\)) among all the tested composites, coupled with the lowest resistivity at room temperature. This suggests that an optimal NBT/TiBaO\(_3\) ratio of 1:3 achieves a desirable balance of low resistivity and high PTC performance, as highlighted in Figure \ref{fig:figure 3} and Table \ref{tab:table2}. To refine the material's thermal characteristics, particularly to achieve a lower Curie temperature, dioctyl phthalate (DOP) was incorporated as a plasticizing agent into the composite. The subsequent investigation, depicted in Figure \ref{fig: figure 4}a, varied the DOP content from 0\% to 20\% while maintaining a consistent carbon black mass fraction of 4\% and an NBT/TiBaO\(_3\) ratio of 1:3. The influence of DOP on the Curie temperature for the composite material labeled A4-E4 is detailed in Table \ref{table: table 3}. The data indicate that an increase in DOP content correlates with a notable reduction in the Curie temperature. This outcome is attributed to DOP's ability to act as a plasticizer, which interferes with the secondary bonding between polymer molecules and weakens the van der Waals forces between molecular chains. This disruption enhances the mobility of the molecular chains, thereby improving the resin's processability and reducing its crystallinity. As DOP concentration rises, the base material's crystal structure becomes more refined, leading to finer crystals that, at higher temperatures, are more prone to melting and decreasing in size. These changes in crystal volume within the conductive channels effectively lower the temperature threshold required to interrupt the conductive pathways, resulting in a decrease in the Curie temperature.
\begin{table}[h]
\centering
\caption{Curie temperature of model A4-E4.}
\begin{tabular}{cccccc}
\toprule
Model Number & A4 & B4 & C4 & D4 & E4 \\
\midrule
Content for DOP & 0 & 5 & 10 & 15 & 20 \\
Curie temperature (°C) & 359.6 & 353.3 & 341.5 & 318.7 & 311.2 \\
\bottomrule
\end{tabular}
\label{table: table 3}
\end{table}

The influence of varying DOP mass fractions ($\omega$) on the room temperature resistivity ($\rho$) and PTC behavior ($P$) of the composite is presented in Figure 4b. The $\rho$-$\omega$ curve distinctly shows that resistivity increases as DOP content is augmented. Conversely, the $P$-$\omega$ curve reveals a gradual reduction in PTC intensity with higher DOP content. When the DOP content is low, DOP crystallization is less likely, resulting in minimal thinning effects. Generally, once the temperature exceeds the Curie point, there is a significant crystalline transition, which substantially alters the material's resistivity, leading to a strong PTC effect. As DOP content increases, the internal crystalline structure of the composite becomes more refined, and when the temperature surpasses the Curie temperature, the crystalline phase undergoes only minor changes, which reduces the peak resistivity. Moreover, the resistivity at room temperature also rises as DOP content increases. As a result, the PTC intensity diminishes with higher DOP levels. According to Figure \ref{fig: figure 4}, a DOP content of 5\% is found to be optimal, yielding low room temperature resistivity and high PTC strength in the composite.

\subsection{Repeat experiment of the Positive temperature coefficient effect for model 4C}\label{sec: 3.2}
\begin{figure}[h!]
    \centering
    \begin{subfigure}[b]{0.5\textwidth}
        \centering
        \includegraphics[width=\textwidth]{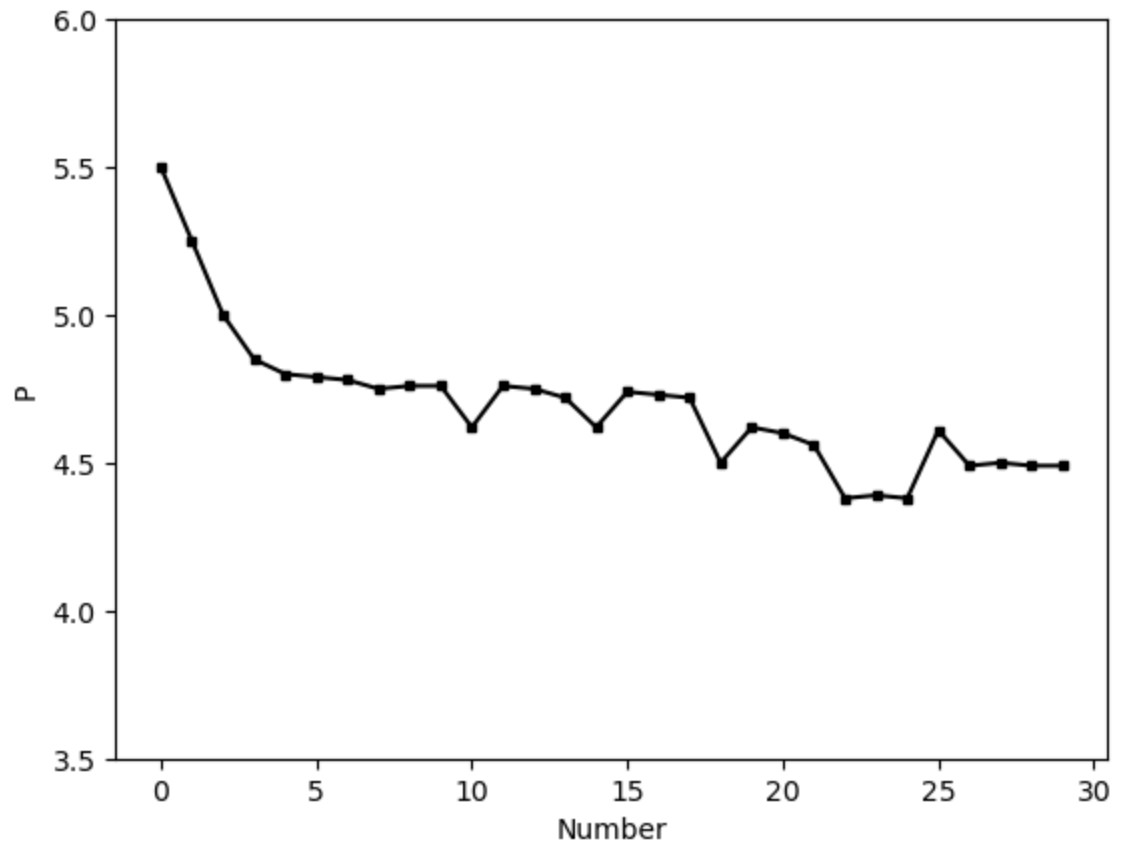}
        \label{fig:figure5a}
    \end{subfigure}
    \hfill
    \begin{subfigure}[b]{0.49\textwidth}
        \centering
        \includegraphics[width=\textwidth]{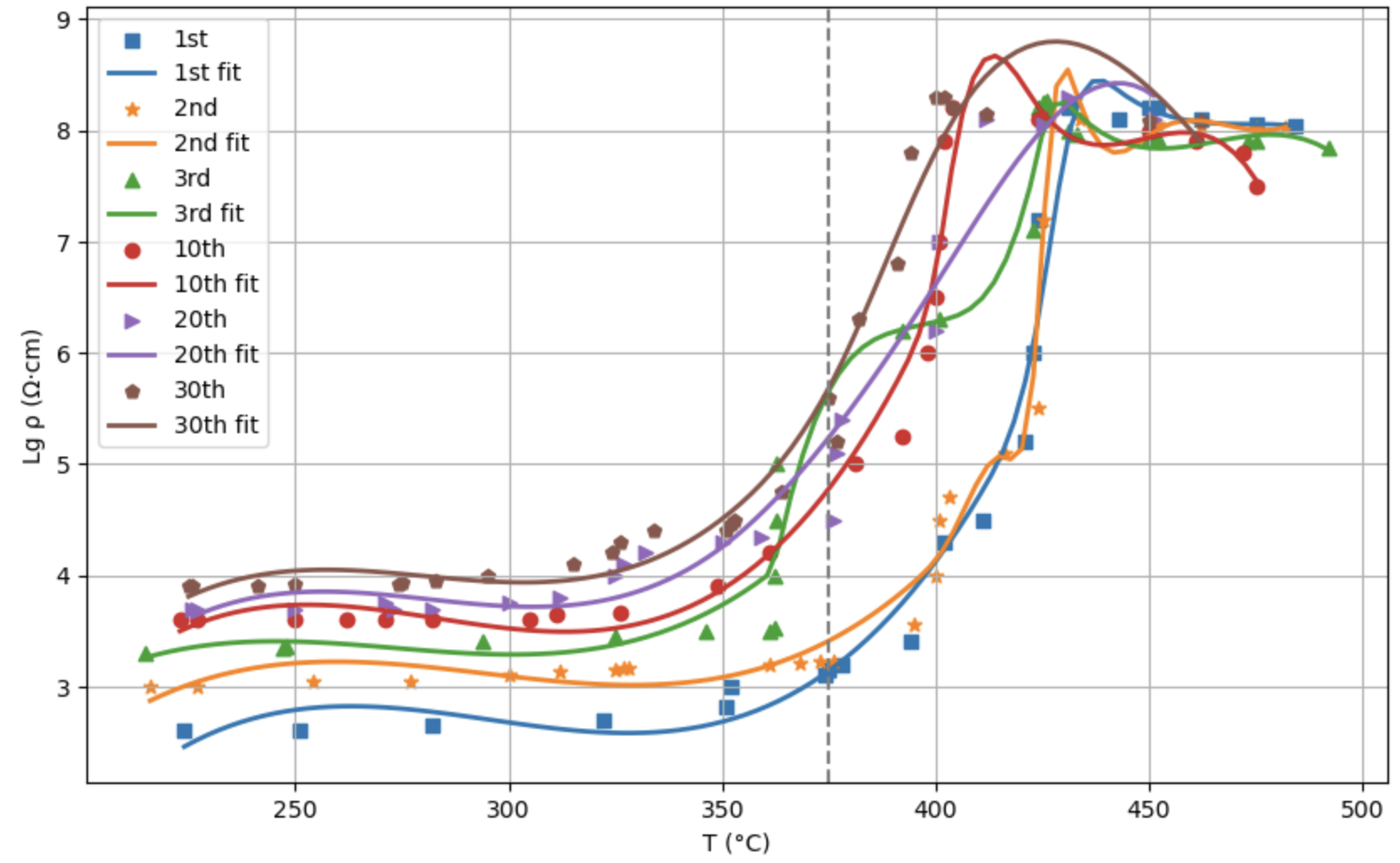}
        \label{fig:figure5b}
    \end{subfigure}
    \caption{Impact of Dioctyl Phthalate (DOP) concentration on material characteristics. (a) Logarithmic analysis of material properties across varying DOP concentrations. (b) Relationship between PTC intensity and resistivity under different DOP contents.}
    \label{fig: figure 5}
\end{figure}

To assess the reproducibility of PTC behavior at the Curie temperature, the material designated as sample 4C, recognized for its optimal properties, was selected as the subject of this study. The sample was subjected to 30 thermal cycles, with the temperature range spanning from 200°C to 500°C. The relationship between PTC strength and the number of thermal cycles is presented in Figure \ref{fig: figure 5}a, highlighting the material's stability under repeated thermal stress.

The analysis demonstrates a significant decline in the PTC intensity of the composite material as the number of thermal cycles increases. Notably, after four cycles, the degree of variation in PTC intensity lessens, although the material still maintains strong PTC characteristics. This trend is depicted in the logarithmic resistivity-temperature curve shown in Figure \ref{fig: figure 5}b. Although resistivity remains relatively constant at elevated temperatures (400°C to 500°C), an increase in room temperature resistivity is evident, which plays a crucial role in the observed reduction of PTC intensity. This increase in room temperature resistivity can be attributed to reversible modifications in the composite’s microstructure, such as changes in volume, aggregation state, and the distribution of conductive particles that occur during thermal cycling. These microstructural changes are likely inherent to the material itself, as corroborated by previous studies \cite{reference19, reference20, reference21}. For a more comprehensive explanation, refer to Section \ref{sec: 3.4}. Moreover, Figure \ref{fig: figure 5}b reveals that the Curie temperature of sample 4C remains unchanged despite the number of thermal cycles. This stability is due to the phase transition temperature of the composite being an intrinsic property of the material, which renders it unaffected by repeated thermal cycling.

\subsection{Positive temperature coefficient (PTC) Effect at Different NBT Proportions}

The inclusion of NBT in the material matrix significantly enhanced its PTC characteristics, although this was accompanied by an increase in room temperature resistivity, as detailed in Table 4. The temperature coefficients, \(\alpha\), listed in Table 4, were determined using a more intricate expression that incorporates the derivative of resistivity with respect to temperature \cite{reference22}:

\[
\alpha = \frac{1}{T_2 - T_1} \times \frac{d}{dT}\left(\ln\left(\frac{R_2}{R_1}\right)\right)
\]

\begin{figure}[H]
    \centering
    \includegraphics[width=0.6\linewidth]{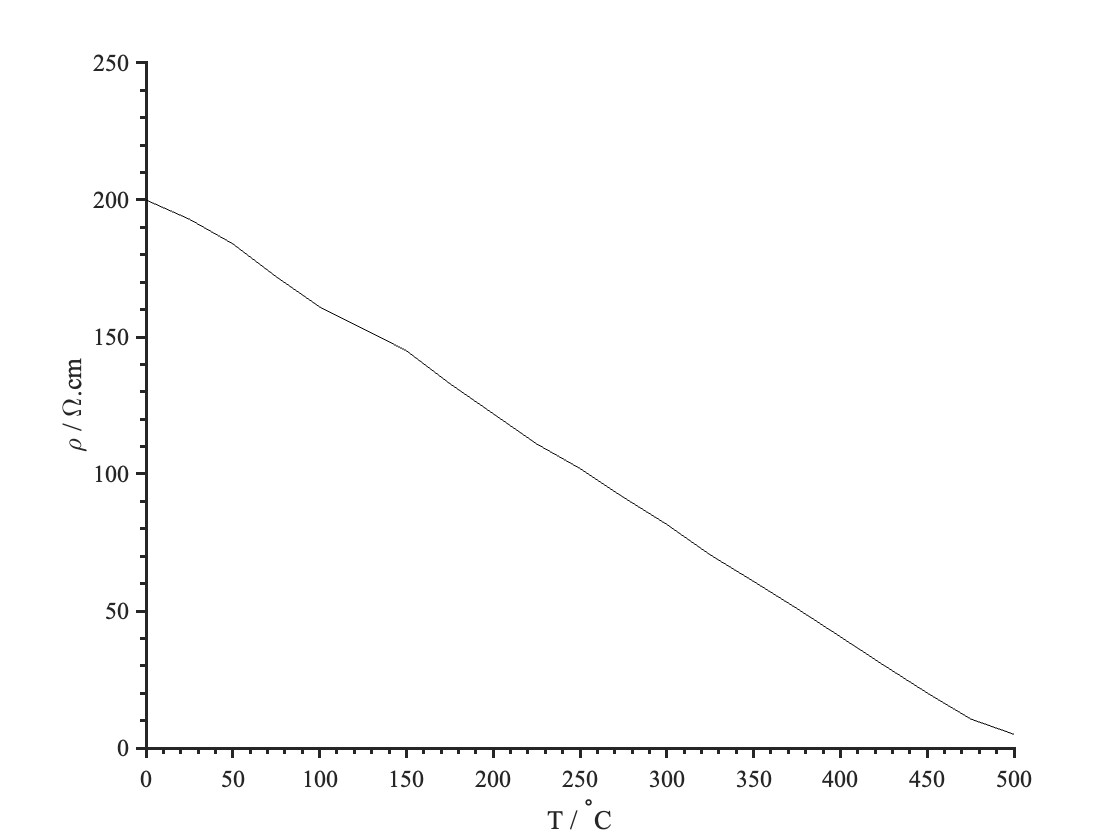}
    \caption{Temperature(°C)-dependent resistivity (\(\rho\)) profiles for samples with varying NBT Proportions}
    \label{fig:Figure 6}
\end{figure}

Here, \(T_1\) denotes the Curie temperature \(T_C\), while \(T_2\) (\(T_2 > T_C\)) corresponds to the temperature at which the derivative \(\frac{d\rho}{dT}\) is evaluated, as depicted in Figure \ref{fig:Figure 6}. The terms \(R_1\) and \(R_2\) represent the resistances at temperatures \(T_1\) and \(T_2\), respectively.
\begin{table}[H]
\centering
\small
\begin{tabular}{>{\centering\arraybackslash}m{3cm} >{\centering\arraybackslash}m{3cm} >{\centering\arraybackslash}m{3cm}}
\toprule
\textbf{NBT proportion (wt\%)} & \textbf{resistivity ($\Omega \cdot$cm)} & \textbf{Resistivity-temperature factor (\% $^\circ$C$^{-1}$)} \\
\midrule
0 & 64.2 & 8.4 \\
0.1 & 98.3 & 9.7 \\
0.5 & 106.36 & 12.6 \\
1 & 153.64 & 17.9 \\
1.5 & 318.45 & 22.4 \\
\bottomrule
\end{tabular}
\caption{Summary of PTC characteristics for samples with varying NBT concentrations}
\label{table:table 4}
\end{table}
\begin{figure}[H]
    \centering
    \includegraphics[width=0.6\linewidth]{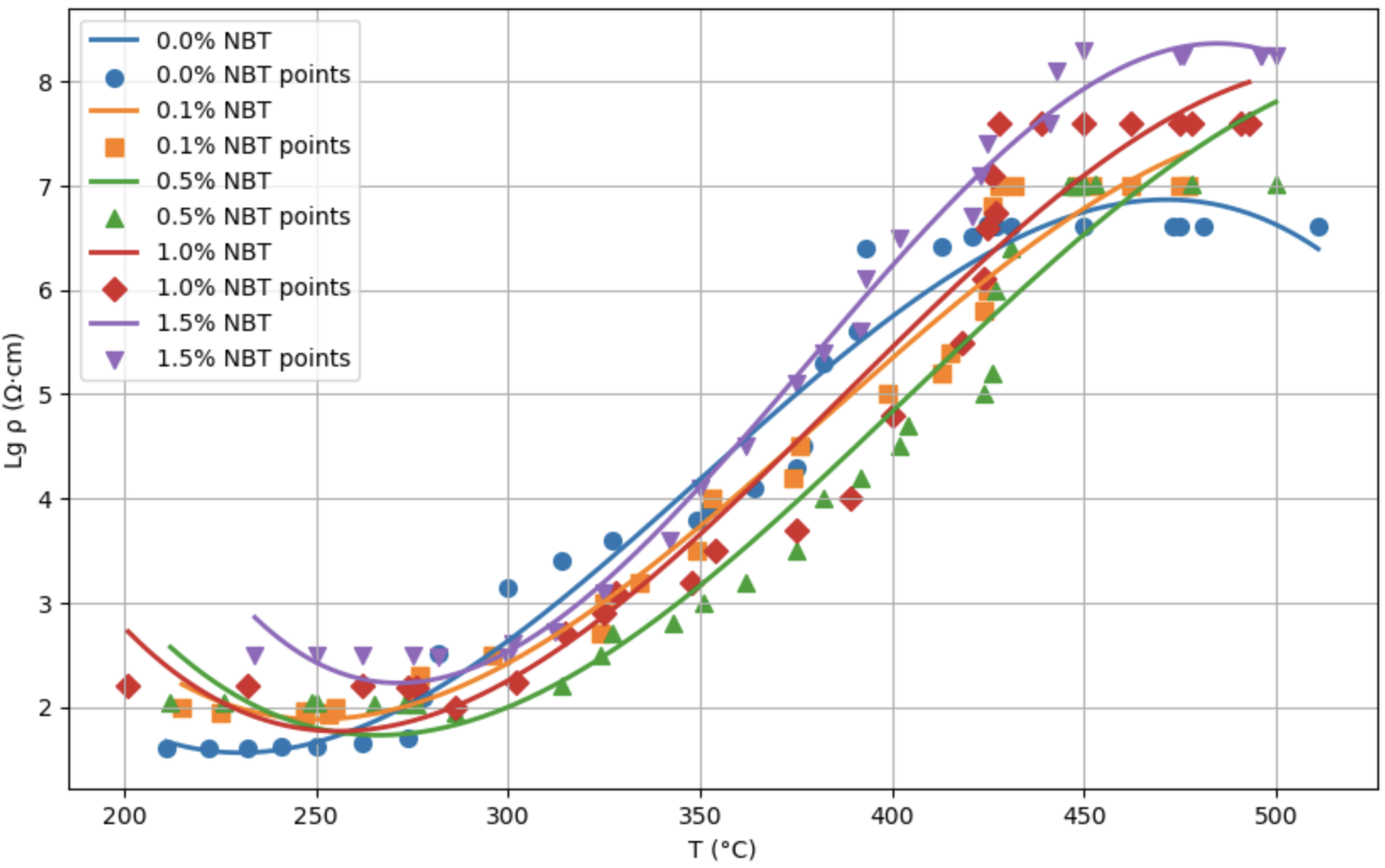}
    \caption{The ρ–T curves of samples with different contents of NBT.}
    \label{fig: figure 7}
\end{figure}

As illustrated in Figure \ref{fig: figure 7}, for the 0.0\% NBT, the initial resistivity is relatively low. Moreover, it increases gradually with the ascending temperature. Around the curing point, the resistivity attains an equilibrium point and keeps increasing, eventually stabilizing at higher temperatures. Secondly, for 0.1\% NBT, in comparison to 0.0\% NBT, the initial resistivity is slightly higher. As the temperature rises, the resistivity increases more rapidly, reaching the equilibrium point before continuing to ascend. The maximum resistivity is higher than that of the 0.0\% NBT material. For 0.5\% NBT, the initial resistivity is higher than the previous two, and it grows rapidly with the increasing temperature. It reaches an equilibrium point and starts to decrease at a slower rate than other percentages of NBT, with the final resistivity being higher than that of the materials with lower NBT content. This suggests that 0.5\% NBT strikes a good balance between the initial resistivity and the resistivity at higher temperatures. For 1.0\% NBT, the initial resistivity further escalates, demonstrating a significant increase with temperature. It reaches an equilibrium point and continues to increase, with the final resistivity conspicuously higher than that of materials with lower NBT content. While providing higher resistivity, it might not be as balanced as the 0.5\% NBT material. Lastly, for 1.5\% NBT which possesses the highest initial resistivity, the resistivity of it surges sharply with the rising temperature. After reaching the equilibrium point, the resistivity continues to rise. What's more, it attains the highest final resistivity, which much higher than other NBT content materials. This indicates that although 1.5\% NBT offers the highest resistivity, it might be excessive for applications where moderate resistivity suffices. Thus, the sample with 0.5\% NBT is ideal for this sample compound due to the balance between the temperature and resistivity in the high-temperature cases.

\subsection{Model Analysis of Microscopic structure}\label{sec: 3.4}
\begin{figure}[H]
    \centering
    \includegraphics[width=0.645\linewidth]{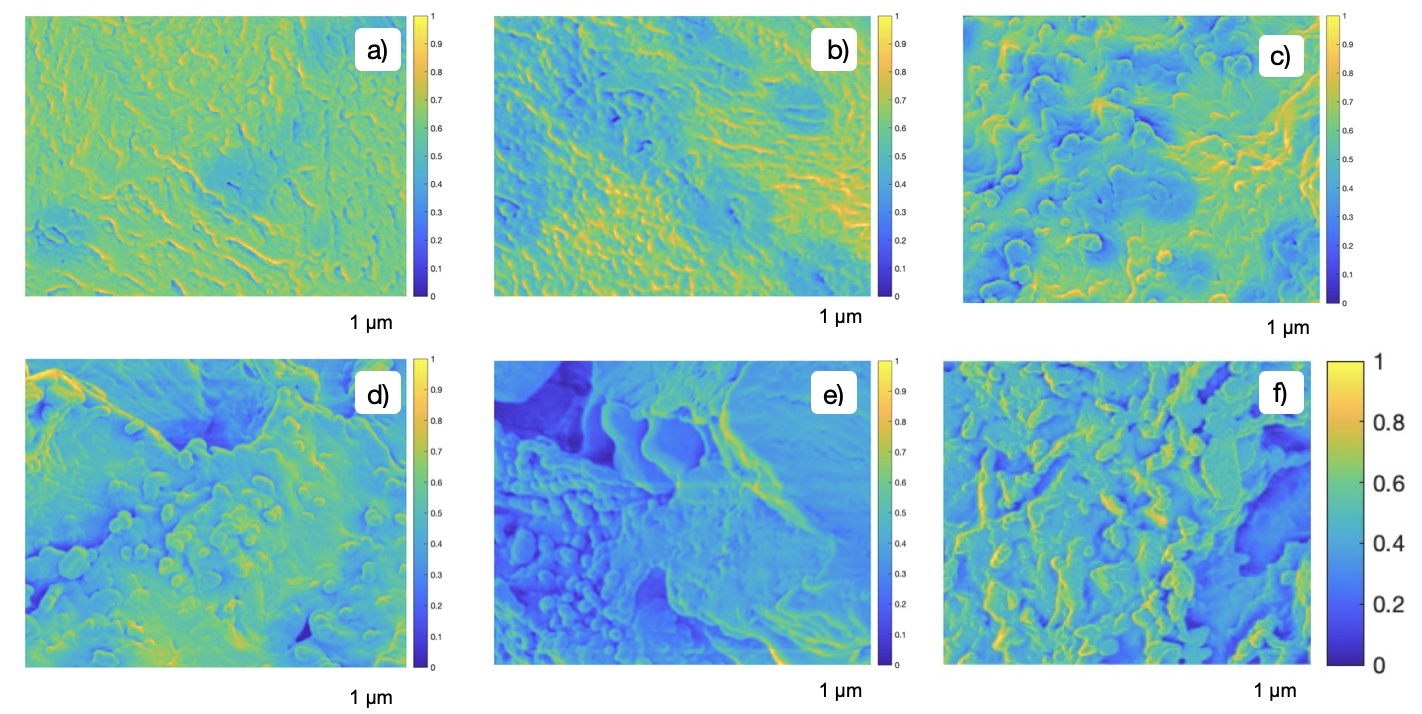}
    \caption{SEM Micrographs of PTC at different CB proportion featuring Varying CB Weight Percentages: (a) 1\%, (b) 2\%, (c) 4\%, (d) 4\%, (e) 7\%, and (f) 10\%.}
    \label{fig:fig 8}
\end{figure}

\begin{figure}[H]
    \centering
    \includegraphics[width=0.7\linewidth]{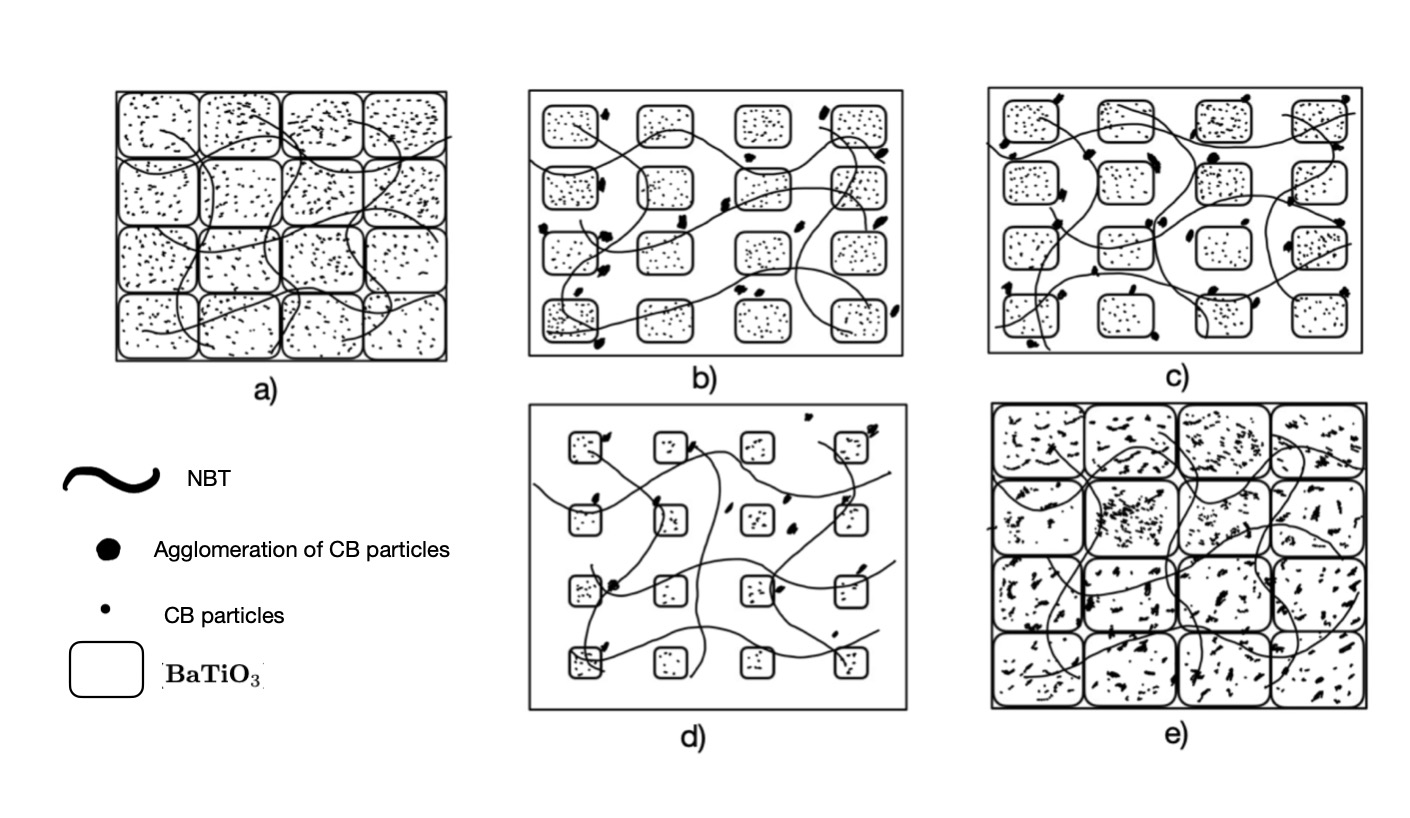}
    \caption{Schematic Representation of Microstructural Evolution (a–e) in PTC Materials Throughout a Thermal Cycle}
    \label{fig:fig 9}
\end{figure}

The SEM micrographs displayed in Figure 8 illustrate how varying the carbon black (CB) content influences the microstructural features of the composites. To better contextualize these observations, refer to the data presented in Figure \ref{fig: figure 2}. As CB concentration varies, the resulting composites exhibit distinct structural characteristics and distribution patterns of CB. For example, in Figures \ref{fig:fig 8}a and \ref{fig:fig 8}b, corresponding to 1\% and 2\% CB, the composites reveal a smooth surface morphology; however, the CB network is discontinuous. This leads to an "Island-Island" pattern of CB dispersion, where the lack of a fully interconnected network results in higher resistivity and consequently lower PTC intensity. When the CB content is increased to 4\%, as shown in Figures \ref{fig:fig 8}c and \ref{fig:fig 8}d, the microstructure changes significantly. The CB particles are more evenly distributed within the NBT/BaTiO\(_3\) matrix, facilitating the development of a continuous conductive pathway that lowers the resistivity at room temperature. Despite this, when the material is heated above the Curie point, the separation between CB regions interrupts the conductive network, which in turn elevates the resistivity and enhances the PTC intensity. Further modifications in CB content to 7\% and 10\%, illustrated in Figures \ref{fig:fig 8}e and \ref{fig:fig 8}f, lead to the formation of a near-complete conductive network. However, the excessive amount of CB also causes agglomeration in certain areas, which paradoxically reduces the overall room temperature resistivity but does not sufficiently disrupt the conductivity at higher temperatures. This results in a modest change in resistivity and a lower-than-expected PTC intensity. To provide a clearer understanding of these phenomena, a schematic of the microstructural transition during thermal cycling is presented in Figure \ref{fig:fig 9}.

The observed PTC behavior can be attributed to the microstructural dynamics within the composite material. Initially, as depicted in the model shown in Figure \ref{fig:fig 9}a, when the temperature is maintained below the Curie point, the CB particles are dispersed randomly throughout the matrix due to the vigorous mixing during preparation. Upon reaching a specific concentration threshold, the conductive particles begin to establish contact, resulting in the formation of a continuous conductive network characterized by high conductivity and low resistivity. As the temperature surpasses the phase transition threshold, the composite's internal structure undergoes significant changes. Figures \ref{fig:fig 9}b and \ref{fig:fig 9}c demonstrate that the crystal volume decreases, leading to the formation of CB agglomerates and an increase in the distance between conductive regions. This disruption in the conductive network causes a notable rise in resistivity. When the temperature is subsequently reduced, as illustrated in Figures \ref{fig:fig 9}d and \ref{fig:fig 9}e, the crystals begin to reform, and the CB agglomerates break down into smaller clusters. However, the material does not fully revert to the well-dispersed state observed in Figure \ref{fig:fig 9}a. Additionally, the extent of carbon black agglomeration is limited, which mitigates the increase in room temperature resistivity after multiple thermal cycles. This model of microstructural transition is instrumental in understanding the gradual increase in room temperature resistivity due to thermal cycling, as further elaborated in Section \ref{sec: 3.2}.

\section{Conclusion}\label{sec:conclusion}

This research presents the triumphant synthesis and comprehensive characterization of a novel positive temperature coefficient (PTC) material featuring a Curie temperature at 353°C, tailored for applications in frigid environments. The composite material, composed of 4 wt\% carbon black (CB), 0.5 wt\% NBT (Na${_{0.5}}$Bi${_{0.5}}$TiO$_3$), and 5 wt\% DOP (dioctyl phthalate) within a BaTiO$_3$ matrix, exhibited an outstanding PTC intensity at 5.8 and resistivity of 590 $\Omega \cdot$cm at room temperature. The optimal equilibrium between the CB content and the NBT/BaTiO$_3$ ratio was pivotal in attaining these properties, with 4\% CB and a 1:3 ratio of NBT to BaTiO$_3$ being identified as optimal. Additionally, the incorporation of 5\% DOP effectively lowered the Curie temperature while maintaining high PTC strength and low resistivity. These discoveries imply that this material is highly appropriate for applications demanding reliable thermal control in cold circumstances, such as the protection and treatment of human hypothermia.

The study also assessed the reproducibility of the material's PTC effect through 30 thermal cycles ranging from 200°C to 500°C. The outcomes demonstrated that although the PTC intensity marginally declined with repeated cycling, mainly attributed to an increase in the resistivity at room temperature, the Curie temperature remained stable, reflecting the inherent thermal stability of the material. SEM analysis disclosed that at the optimum CB content, the conductive network within the composite was evenly distributed, resulting in substantial enhancements in PTC strength and resistivity. The proposed microstructural transition model elucidated the PTC behavior, depicting how the conductive network forms and disrupts with variations in temperature. This comprehension is vital for further enhancing the material's attributes. Overall, the development of this PTC material constitutes a significant progress in the field, presenting new prospects for thermal management in various low-temperature applications, particularly in medical scenarios for treating hypothermia.

\section{Appendix A: Equations for BaTiO$_3$}\label{sec: Appendix}
We start by considering the differential form of the displacement field equation, incorporating the divergence theorem:

\begin{equation}
\oint_{\partial V} \mathbf{D} \cdot d\mathbf{A} = \int_V \nabla \cdot \mathbf{D} \, dV = e\left(2N_d z_0 - N_s\right)
\end{equation}

This can be further expressed as:

\begin{equation}
2z_0 = \frac{e\int_V N_i \, dV - 2\int_{\partial V} P_s \cos \theta \, dA}{e\int_V N_d \, dV} = \frac{\int_V N_e \, dV}{\int_V N_d \, dV}
\end{equation}

It is crucial to note that \(N_e\) varies with temperature due to the dependency of \(P_s\) on \(T\). To normalize \(N_s\), we define a scaling function \(s(T)\) that describes \(N_e(T)\) at a specific reference temperature, such as 25°C:

\begin{equation}
\frac{d}{dT} \left( e\int_V N_e(25^\circ \text{C}) \, dV \right) = 2s(T) \frac{d}{dT} \left( \int_{\partial V} P_s(25^\circ \text{C}) \, dA \right)
\end{equation}

Alternatively, this can be reformulated as:

\begin{equation}
1 + S(T) = \frac{e \int_V N_s \, dV}{2 \int_{\partial V} P_c(25^\circ \text{C}) \, dA}
\end{equation}

For an arbitrary temperature \(T\), the expression becomes:

\begin{equation}
\frac{d}{dT} \left( e\int_V N_e(T) \, dV \right) = 2\left(1 + S(T)\right)\frac{d}{dT} \left( \int_{\partial V} P_s(25^\circ \text{C}) \, dA \right) - \frac{d}{dT} \left( \int_{\partial V} P_s(T) \, dA \right)
\end{equation}

When \(T\) exceeds the Curie point and \(P_s\) approaches zero, this simplifies to:

\begin{equation}
\frac{d}{dT} \left( e\int_V N_e(T) \, dV \right) = 2\left(1 + S(T)\right)\frac{d}{dT} \left( \int_{\partial V} P_s(25^\circ \text{C}) \, dA \right), \quad T > T_c
\end{equation}

The free energy density \(F\) of the crystal in an unclamped state can be generalized by incorporating higher-order terms:

\begin{equation}
F = \frac{1}{2}\alpha P^2 + \frac{1}{4}\beta P^4 + \frac{1}{6}\gamma P^6 + \frac{1}{8}\delta P^8 + \cdots
\end{equation}

From this, the electric field \(E\) along the axis, related to the free energy density, is given by the derivative:

\begin{equation}
E = \frac{\partial F}{\partial P} = \alpha P + \beta P^3 + \gamma P^5 + \delta P^7 + \cdots
\end{equation}

The condition defining the Curie temperature \(T_c\) is modified to:

\begin{equation}
\frac{\alpha \gamma}{\beta^2} \times \delta = \frac{3}{16}
\end{equation}

For spontaneous polarization \(P_s\), we introduce a piecewise function dependent on temperature:

\begin{equation}
P_s = 
\begin{cases} 
\left(\frac{\beta}{2\gamma}\right)^{\frac{1}{3}} \left[1 + \left(1 - \frac{4\alpha \gamma}{\beta^2}\right)^{\frac{1}{4}}\right]^{\frac{1}{3}} & \text{for } T \leq T_c \\
0 & \text{for } T > T_c 
\end{cases}
\end{equation}

The displacement field \( \mathbf{D} \) is expressed as a function of both electric field \( \mathbf{E} \) and polarization \( \mathbf{P} \):

\begin{equation}
\mathbf{D} = \epsilon_0 \mathbf{E} + \mathbf{P}
\end{equation}

To approximate the relative permittivity, the following relationship is employed:

\begin{equation}
\frac{dD}{dE} = \epsilon_0 + \frac{dP}{dE} \approx \frac{dP}{dE} 
\end{equation}

A significant depletion-inversion layer within the material can be defined by examining the spatial derivative of the electric field:

\begin{equation}
\epsilon_0 \frac{d^2 E}{dz^2} = \rho = eN_d
\end{equation}

This approximation holds true for interface potentials within the range \(2kT \ll -e\phi_0 \ll E_g - \Delta E\), where \(\Delta E = 2kT \ln \left( \frac{N_C}{N_V} \right)\) and \(E_g\) denotes the energy gap. By applying Gauss' law, \(\nabla \cdot \mathbf{D} = \rho\), the corresponding differential equation simplifies to:

\begin{equation}
\frac{d^2 P}{dz^2} = eN_d
\end{equation}

Now, by combining equations 8 and 13, we arrive at the following differential equation:

\begin{equation}
eN_d \frac{d\phi}{dP} = \alpha P + \beta P^3 + \gamma P^5
\end{equation}

Integrating with respect to \( P \) yields:

\begin{equation}
-eN_d\phi = \frac{1}{2}\alpha (P - P_s)^2 + \frac{1}{4}\beta (P - P_s)^4 + \frac{1}{6}\gamma (P - P_s)^6
\end{equation}

The difference \( P - P_s \) can be determined by integrating equation 13 from \( z_0 \) to \( z \):

\begin{equation}
P - P_s = \int_{z_0}^{z} eN_d \, dz = eN_d(z - z_0)
\end{equation}

Substituting \( z_0 \) from equation 2, we obtain:

\begin{equation}
-eN_d\phi = \frac{1}{2}\alpha e^2 \left(N_d(z - z_0) - \frac{1}{2}N_e\right)^2 + \frac{1}{4}\beta e^4 \left(N_dz - \frac{1}{2}N_e\right)^4 + \frac{1}{6}\gamma e^6 \left(N_dz - \frac{1}{2}N_e\right)^6
\end{equation}

For isotropic materials, where \( \alpha = \frac{1}{\epsilon \epsilon_0} \), equation 21 modifies as follows:

\begin{equation}
\left\{
\begin{aligned}
    \frac{\phi}{\phi_0} &= \frac{(z - z_0)^2}{z_0^2} \\
    -\phi_0 &= \frac{eN_s^2}{8\epsilon \epsilon_0 N_d}
\end{aligned}
\right.
\end{equation}

The normalized resistivity \( R_0 \) is then given by:

\begin{equation}
R_0 = \lim_{v \to 0} R = (e\mu N_d)^{-1} \int_{-z_0}^{z_0} \exp[u(z)] \, dz
\end{equation}

The apparent resistivity associated with \( \frac{1}{\delta} \) barriers per meter is \( \frac{R_0}{\delta} \), so for a series combination of grains and boundaries, we have:

\begin{equation}
\kappa = (e\mu N_d)^{-1} + \frac{R_0}{\delta}
\end{equation}

Finally, we express the dimensionless quantity \( \eta \) as:

\begin{equation}
\eta = \kappa e \mu N_d = 1 + \frac{1}{\delta} \int_{-z_0}^{z_0} \exp \left[u(z)\right] \, dz
\end{equation}

\subsection{Appendix B: The Refined Electrical Conductivity Model of TiBaO$_3$}
We commence by examining the Gibbs free energy associated with point defects, which can be expressed in terms of the changes in both energy and chemical potential as follows:
\begin{equation}
\Delta G_f = (G_{\text{def}} - G_{\text{host}}) - \int_0^n \left(\frac{\partial G}{\partial n_i}\right)_{\mu_i} \, dn_i + q\left(\mu_e + \int_{V_0}^{V} \frac{\partial E_v}{\partial V} \, dV\right)
\end{equation}
In this expression, \(G_{\text{def}}\) and \(G_{\text{host}}\) denote the Gibbs free energies of the defective and pristine systems, respectively. The term \(\frac{\partial G}{\partial n_i}\) represents the change in the number of atoms for element \(i\) with respect to its chemical potential, \(\mu_i\) is the chemical potential of element \(i\), \(E_v\) is the energy position of the valence band maximum, and \(q\) symbolizes the charge state of the defect.

The chemical potential \(\mu_i\) can be decomposed into its ground state value and a deviation as a function of state variables:
\begin{equation}
    \mu_i = \mu_i^0 + \delta \mu_i = \mu_i^0 + \left(\frac{\partial \mu_i}{\partial T}\right)_P dT + \left(\frac{\partial \mu_i}{\partial P}\right)_T dP
\end{equation}

The concentration of defects as a function of temperature can be described by the following integral:
\begin{equation}
    c_i = \int_0^{T} N_i(T') \exp \left( -\frac{\Delta G_f(T')}{k_B T'} \right) \, dT'
\end{equation}
where \(N_i(T')\) denotes the temperature-dependent number of available sites per unit volume for the defect. The charge neutrality condition within the system involves both sums and integrals over charge distributions:
\begin{equation}
    n_e + \sum_k \int_0^V q_k \, dV = n_h + \sum_j \int_0^V q_j \, dV
\end{equation}
where \(n_e\) and \(n_h\) are the electron and hole concentrations, respectively, and \(\sum_k q_k\) and \(\sum_j q_j\) represent the charge concentrations from donor and acceptor defects. The intrinsic electron and hole concentrations can be derived by integrating the density of states over energy:
\begin{equation}
    n_e = \int_{\epsilon_c}^{\infty} \int_{0}^{E} g_c(E') f(E', \mu_e) \, dE' \, dE
\end{equation}
\begin{equation}
    n_h = \int_{-\infty}^{\epsilon_v} \int_{E}^{0} g_v(E') \left[1 - f(E', \mu_e)\right] \, dE' \, dE
\end{equation}
where \(g_c(E)\) and \(g_v(E)\) are the density of states functions for the conduction and valence bands, respectively, and \(f(E, \mu_e)\) is the Fermi-Dirac distribution function.

The carrier concentration from extrinsic donor and acceptor defects can be described by:
\begin{equation}
    n_D^{\text{ext}} = N_D^{\text{ext}} \left[ 1 - f(E_g - E_D, \mu_e) \right] = \int_{0}^{E_D} N_D^{\text{ext}}(E') \, dE'
\end{equation}

\begin{equation}
    n_A^{\text{ext}} = \int_{E_A}^{\infty} N_A^{\text{ext}}(E') f(E', \mu_e) \, dE'
\end{equation}

The total Gibbs free energy change of the system can be approximated by incorporating configurational entropy:
\begin{equation}
    \Delta G_{\text{total}} = \int_0^N \left( \sum_i \Delta G_f \, dN_i - k_B T \ln \Omega \, dN_i \right)
\end{equation}
where \(N_i\) represents the number of defects, and \(\Omega\) denotes the total number of possible configurations. The total electrical conductivity \(\sigma\) is given by the summation of integrals over contributions from all charge carriers:
\begin{equation}
    \sigma = \sum_i \int_0^V \mu_i(V') c_i(V') q_i e \, dV'
\end{equation}
where \(\mu_i(V')\) denotes the position-dependent mobility, \(c_i(V')\) represents the concentration, and \(q_i\) is the charge number.

The contributions to electrical conductivity from electrons and holes involve integrals over their respective distributions:
\begin{equation}
    \sigma_e = \int_0^V \mu_e(V') n_e(V') e \, dV'
\end{equation}

\begin{equation}
    \sigma_h = \int_0^V \mu_h(V') n_h(V') e \, dV'
\end{equation}

The electron mobility can be described as an integral over temperature:
\begin{equation}
    \mu_e = \int_{0}^{T} \frac{A(T')}{(T')^{3/2}} \exp \left( -\frac{E_a}{k_B T'} \right) \, dT'
\end{equation}
where \(A(T')\) and \(E_a\) are empirical fitting parameters. The ionic conductivity is similarly described by:
\begin{equation}
    \sigma_{\text{ion}} = \sum_i \int_0^V \frac{q_i^2 e^2 D_i(V') c_i(V')}{k_B T} \, dV'
\end{equation}
where \(D_i(V')\) represents the position-dependent diffusion coefficient of the defect.

The chemical potentials of the constituent elements are constrained by the following integral condition:
\begin{equation}
   \delta \mu_{\text{Ba}} + \delta \mu_{\text{Ti}} + 3 \delta \mu_{\text{O}} = \int_{0}^{V} \Delta H_f[\text{BaTiO}_3(V')] \, dV'
\end{equation}

The formation conditions for competing phases can be expressed as:
\begin{equation}
    \delta \mu_{\text{Ba}} + \delta \mu_{\text{O}} \leq \int_{0}^{V} \Delta H_f[\text{BaO}(V')] \, dV'
\end{equation}

\begin{equation}
   \delta \mu_{\text{Ti}} + 2 \delta \mu_{\text{O}} \leq \int_{0}^{V} \Delta H_f[\text{TiO}_2(V')] \, dV'
\end{equation}

\section*{Declaration of Conflict of Interest}
The author(s) declare that there is no conflict of interest regarding the publication of this paper.
\end{document}